\documentclass[twocolumn,pra,showpacs,preprintnumbers,amsmath,amssymb]{revtex4}

\usepackage{bm}

\begin{document}

\title{No spin-statistics connection in nonrelativistic quantum mechanics}

\author{R. E. Allen and A. R. Mondragon}

\affiliation{Center for Theoretical Physics, Texas A\&M University, 
College Station, Texas 77843}

\date{\today}

\begin{abstract}
We emphasize that there is no spin-statistics connection in nonrelativistic 
quantum mechanics. In several recent papers, including Phys. Rev. A {\bf 67}, 
042102 (2003), quantum mechanics is modified so as to force a spin-statistics 
connection, but the resulting theory is quite different from standard physics.
\end{abstract}


\maketitle

It has been known for many years that there is a spin-statistics connection 
in relativistic quantum field theory~[1-4] but not in nonrelativistic quantum
mechanics~\cite{wightman2}. However, several recent papers~[6-8] have 
led to some confusion regarding the second point. 

Let us first remind ourselves why there is no spin-statistics
theorem in nonrelativistic quantum mechanics. The essential reason is that
the restrictions that imply a spin-statistics connection in relativistic
field theory are no longer meaningful in nonrelativistic physics. For
example, Weinberg's textbook~\cite{weinberg} provides a relatively simple and
physical proof based on microcausality, or the requirement that commutators
associated with observable quantities vanish for spacelike separations. In
nonrelativistic physics, causality is still a meaningful requirement, but
microcausality is not, because there is no longer a light cone. This proof
then does not apply in the nonrelativistic case, and the same is true of the
other proofs based on Lorentz invariance.

There are nonrelativistic wavefunctions for either $N$ fermions 
or $N$ bosons with any spin (0, 1/2, 1, 3/2, ...). For example, a basis 
function with the form 
\begin{equation}
\Psi \left( \mathbf{r}_{1},\mathbf{r}_{2}\right) =\left( \phi _{1}\left( 
\mathbf{r}_{1}\right) \phi _{2}\left( \mathbf{r}_{2}\right) -\phi _{1}\left( 
\mathbf{r}_{2}\right) \phi _{2}\left( \mathbf{r}_{1}\right) \right) /\sqrt{2}
\end{equation}
is acceptable for spin-zero fermions, where $\phi $ is a simple scalar. More
generally, a basis function with the form 
\begin{eqnarray}
\Psi \left( \mathbf{r}_{1},\mathbf{r}_{2},...,\mathbf{r}_{N}\right) &=&
\mathcal{A\,}\prod_{i=1}^{N}\psi _{i}\left( \mathbf{r}_{i}\right)
\mbox{  ,  fermions} \\
&=&\mathcal{S\,}\prod_{i=1}^{N}\psi _{i}\left( \mathbf{r}_{i}\right) 
\mbox{  ,  bosons}
\end{eqnarray}
is appropriate for $N$ particles with any spin. Here $\mathcal{A}$ or 
$\mathcal{S}$ represents antisymmetrization or symmetrization of the 
product (with insertion of the correct normalization factor). 
Each $\psi $ is a function corresponding to the desired spin $s$; for
example, $\psi $ is a $2$-component spinor if $s=1/2$. A nonrelativistic
field theory can then be constructed in the usual way, having fermions or
bosons with any spin~\cite{fetter}. The field operator consistently 
transforms as both a field and a quantum operator~\cite{wightman2}.

According to Ref. 6, on the other hand, (1) is not an acceptable
wavefunction. This conclusion was reached because quantum physics was
modified by adding an unusual constraint: In the words of Ref. 6, ``The
approach used here is based on the requirement that the point $\{\mathbf{r}
_{1},\mathbf{r}_{2}\}$ in the configuration space for two identical spinless
particles is the same point as $\{\mathbf{r}_{2},\mathbf{r}_{1}\}$.'' But
this requirement implies that the wavefunction must return to its original
value when $\left( \mathbf{r}_{1},\mathbf{r}_{2}\right) $ is transformed to 
$\left( \mathbf{r}_{2},\mathbf{r}_{1}\right) $: 
\begin{equation}
\Psi \left( \mathbf{r}_{2},\mathbf{r}_{1}\right) =\Psi \left( \mathbf{r}_{1},
\mathbf{r}_{2}\right) .
\end{equation}
I.e., the two-particle wavefunction is only allowed to acquire the $+$ sign
appropriate for bosons, and is forbidden to acquire the $-$ sign appropriate
for fermions. It is this requirement that forbids spin-zero fermions with 
the wavefunction (1). In Ref. 6, therefore, the spin-statistics connection 
is simply imposed by fiat.

Essentially the same philosophy was used in Refs. 7 and 8. In the words of
Ref. 7, ``we must identify the points $\mathbf{r}$ and $-\mathbf{r}$, since
these correspond to complete interchange of the particles (positions and
spins) and so are indistinguishable.'' (Here $\mathbf{r=r}_{2}-\mathbf{r}
_{1} $ is the relative coordinate.) They then conclude that 
\begin{equation}
|\Psi \left( -\mathbf{r}\right) \rangle =|\Psi \left( \mathbf{r}\right)
\rangle
\end{equation}
where $|\Psi \left( \mathbf{r}\right) \rangle $ specifies the state of the
two particles. ($|\Psi \left( \mathbf{r}\right) \rangle $ is a $\left(
2s+1\right) ^{2}$ dimensional vector, since there are $\left( 2s+1\right) $
components for a single particle.)

With a standard basis for the spins, (5) is untenable, so further 
alteration of quantum theory is required. The authors of Ref. 7 write the 
state of the two particles as 
\begin{equation}
|\Psi \left( \mathbf{r}\right) \rangle =\sum_{M}\Psi _{M}\left( \mathbf{r}
\right) \,|M\left( \mathbf{r}\right) \rangle 
\end{equation}
where the spin basis functions are not standard, but instead are modified to
have a position dependence. This ultimately implies that 
\begin{equation}
\pm \Psi \left( \mathbf{r}\right) \,=\left( -1\right) ^{2s}\Psi 
\left( \mathbf{r}\right) 
\end{equation}
where $\Psi$ is the vector with components $\Psi_{M}$. 
The upper sign holds for bosons and the lower sign for fermions. (See
(3.4) of Ref. 7 and the discussion below this equation.) Then $2s$ must be
even for bosons and odd for fermions. However, this result can be traced back 
to the assumption (5). Again, in the simplest case $s=0$, fermions have 
clearly been banished at the outset.

If one does not impose the unusual constraint (4) or (5), nonrelativistic 
bosons are allowed to have any spin (0, 1/2, 1, 3/2, ...) and the same is 
true of nonrelativistic fermions.

\end{document}